\documentstyle[preprint,tighten,aps,amstex,graphicx,times,floats]{revtex}

\begin{document}

\thispagestyle{empty}

% macros for marking changes
\marginparwidth 1.cm
\setlength{\hoffset}{-1cm}
\newcommand{\mpar}[1]{{\marginpar{\hbadness10000%
                      \sloppy\hfuzz10pt\boldmath\bf\footnotesize#1}}%
                      \typeout{marginpar: #1}\ignorespaces}
\def\mda{\mpar{\hfil$\downarrow$\hfil}\ignorespaces}
\def\mua{\mpar{\hfil$\uparrow$\hfil}\ignorespaces}
\def\mla{\marginpar[\boldmath\hfil$\rightarrow$\hfil]%
                   {\boldmath\hfil$\leftarrow $\hfil}%
                    \typeout{marginpar: $\leftrightarrow$}\ignorespaces}

\renewcommand{\abstractname}{Abstract}
\renewcommand{\figurename}{Figure}
\renewcommand{\refname}{Bibliography}

% peter's conventions 
\newcommand{\eg}{{\it e.g.}\;}
\newcommand{\ie}{{\it i.e.}\;}
\newcommand{\etal}{{\it et al.}\;}
\newcommand{\ibid}{{\it ibid.}\;}

% additional commands 
\newcommand{\mx}{M_{\rm SUSY}}
\newcommand{\pt}{p_{\rm T}}
\newcommand{\et}{E_{\rm T}}
\newcommand{\del}{\varepsilon}
\newcommand{\sla}[1]{/\!\!\!#1}

% some journals 
\newcommand{\zpc}[3]{${\rm Z. Phys.}$ {\bf C#1} (19#2) #3}
\newcommand{\npb}[3]{${\rm Nucl. Phys.}$ {\bf B#1} (19#2)~#3}
\newcommand{\plb}[3]{${\rm Phys. Lett.}$ {\bf B#1} (19#2) #3}
\renewcommand{\prd}[3]{${\rm Phys. Rev.}$ {\bf D#1} (19#2) #3}
\renewcommand{\prl}[3]{${\rm Phys. Rev. Lett.}$ {\bf #1} (19#2) #3}
\newcommand{\prep}[3]{${\rm Phys. Rep.}$ {\bf #1} (19#2) #3}
\newcommand{\fp}[3]{${\rm Fortschr. Phys.}$ {\bf #1} (19#2) #3}
\newcommand{\nc}[3]{${\rm Nuovo Cimento}$ {\bf #1} (19#2) #3}
\newcommand{\ijmp}[3]{${\rm Int. J. Mod. Phys.}$ {\bf #1} (19#2) #3}
\renewcommand{\jcp}[3]{${\rm J. Comp. Phys.}$ {\bf #1} (19#2) #3}
\newcommand{\ptp}[3]{${\rm Prog. Theo. Phys.}$ {\bf #1} (19#2) #3}
\newcommand{\sjnp}[3]{${\rm Sov. J. Nucl. Phys.}$ {\bf #1} (19#2) #3}
\newcommand{\cpc}[3]{${\rm Comp. Phys. Commun.}$ {\bf #1} (19#2) #3}
\newcommand{\mpl}[3]{${\rm Mod. Phys. Lett.}$ {\bf #1} (19#2) #3}
\newcommand{\cmp}[3]{${\rm Commun. Math. Phys.}$ {\bf #1} (19#2) #3}
\newcommand{\jmp}[3]{${\rm J. Math. Phys.}$ {\bf #1} (19#2) #3}
\newcommand{\nim}[3]{${\rm Nucl. Instr. Meth.}$ {\bf #1} (19#2) #3}
\newcommand{\prev}[3]{${\rm Phys. Rev.}$ {\bf #1} (19#2) #3}
\newcommand{\el}[3]{${\rm Europhysics Letters}$ {\bf #1} (19#2) #3}
\renewcommand{\ap}[3]{${\rm Ann. of~Phys.}$ {\bf #1} (19#2) #3}
\newcommand{\jhep}[3]{${\rm JHEP}$ {\bf #1} (19#2) #3}
\newcommand{\jetp}[3]{${\rm JETP}$ {\bf #1} (19#2) #3}
\newcommand{\jetpl}[3]{${\rm JETP Lett.}$ {\bf #1} (19#2) #3}
\newcommand{\acpp}[3]{${\rm Acta Physica Polonica}$ {\bf #1} (19#2) #3}
\newcommand{\science}[3]{${\rm Science}$ {\bf #1} (19#2) #3}
\newcommand{\vj}[4]{${\rm #1~}$ {\bf #2} (19#3) #4}
\newcommand{\ej}[3]{${\bf #1}$ (19#2) #3}
\newcommand{\vjs}[2]{${\rm #1~}$ {\bf #2}}
\newcommand{\hep}[1]{${\tt hep\!-\!ph/}$ {#1}}
\newcommand{\hex}[1]{${\tt hep\!-\!ex/}$ {#1}}
\newcommand{\desy}[1]{${\rm DESY-}${#1}}
\newcommand{\cern}[2]{${\rm CERN-TH}${#1}/{#2}}

\preprint{
\font\fortssbx=cmssbx10 scaled \magstep2
\hbox to \hsize{
% Next 3 lines for UW output only:
% \special{psfile=/NextLibrary/TeX/tex/inputs/uwlogo.ps
%                             hscale=8000 vscale=8000
%                              hoffset=-12 voffset=-2}
\hskip.5in \raise.1in\hbox{\fortssbx University of Wisconsin - Madison}
\hfill\vtop{\hbox{\bf MADPH-99-1101}
            \hbox{February 1999}} }
}

\title{ 
Probing the MSSM Higgs Sector via Weak Boson Fusion at the LHC 
} 

\author{
Tilman Plehn, David Rainwater, and Dieter Zeppenfeld 
} 

\address{ 
Department of Physics, University of Wisconsin, Madison, WI 53706 
} 

\maketitle 

\begin{abstract}
In the MSSM weak boson fusion produces the two CP even Higgs bosons
with a combined strength equivalent to the production of the Standard
Model Higgs boson. The $\tau\tau$ decay mode --- supplemented by
$\gamma\gamma$ --- provides a highly significant signal for at least
one of the CP even Higgs bosons at the LHC with reasonable
luminosity. The accessible parameter space covers the entire physical
range which will be left unexplored by LEP2.
\end{abstract} 

%\pacs{PACS numbers: 13.85.-t, 14.80.Bn, 14.60.Fg} 
% these PACS were chosen for ph1057 %

%%%%%%%%%%%%%%%%%%%%%%%%%%%%%%%  MAIN TEXT  %%%%%%%%%%%%%%%%%%%%%%%%%%%%
\section{Introduction}\label{sec:one}

The search for the Higgs boson and the origin of spontaneous breaking
of the electroweak gauge symmetry is one of the main tasks of the CERN
Large Hadron Collider (LHC). Within the Standard Model (SM), a combination of
search strategies will allow a positive identification of the Higgs
signal~\cite{review}: for small masses ($m_H\lesssim 140$~GeV) the
Higgs boson can be seen as a narrow resonance in inclusive two-photon
events and in associated production in the $t\bar tH$, $b\bar bH$ and
$WH$ channels with subsequent decay
$H\to\gamma\gamma$~\cite{ttH,bbH,WH}. For large Higgs masses
($m_H\gtrsim 130$~GeV), the search in $H\to ZZ^{(*)}\to 4\ell$ events
is promising. Additional modes have been suggested recently: the
inclusive search for $H\to WW^*\to \ell\ell\sla \pt$~\cite{dd}, and
the search for $H\to \gamma\gamma$ or $\tau\tau$ in weak boson fusion
events~\cite{rz,rzh}. With its two forward quark jets, the weak boson
fusion possesses unique characteristics which allow identification
with a very low level of background at the LHC. At the same time,
reconstruction of the $\tau\tau$ invariant mass is possible;
modest luminosity, of order of 30~fb$^{-1}$, should suffice for a
$5\sigma$ signal.\smallskip

In the minimal supersymmetric extension of the SM the
situation is less clear~\cite{review}. The search is open for two CP
even mass eigenstates, $h$ and $H$, for a CP odd $A$, and for a
charged Higgs boson $H^{\pm}$. For large $\tan\beta$, the light
neutral Higgs boson may couple much more strongly to the $T_3=-1/2$
members of the weak isospin doublets than its SM
analogue. As a result, the total width can increase significantly
compared to a SM Higgs boson of the same mass. This comes
at the expense of the branching ratio $B(h\to\gamma\gamma)$, the
cleanest Higgs discovery mode, possibly rendering it unobservable and
forcing the consideration of alternative search channels. Even when
discovery in the inclusive $\gamma\gamma$ channel is possible,
observation in alternative production and decay channels is needed to
measure the various couplings of the Higgs resonance and thus identify
the structure of the Higgs sector~\cite{selfcoup}.\smallskip

In this Letter we explore the reach of weak boson fusion with
subsequent decay to $\tau\tau$ for Higgs bosons in the MSSM
framework. We will show that, except for the low $\tan\beta$ region
which is being excluded by LEP2, the weak boson fusion channels are
most likely to produce significant $h$ and/or $H$ signals.

% ==================================================================
\section{Neutral Higgs Bosons in the MSSM}\label{sec:two}

Some relevant features of the minimal supersymmetric Higgs sector can
be illustrated in a particularly simple approximation~\cite{easy}:
including the leading contributions with respect to $G_F$ and the top
flavor Yukawa coupling, $h_t =m_t/(vs_\beta)$. The qualitative
features remain unchanged in a more detailed description. All our
numerical evaluations make use of a renormalization group improved
next-to-leading order calculation~\cite{one_loop,hdecay}. The
inclusion of two loop effects is not expected to change the results
dramatically~\cite{two_loop}. Including the leading contributions with
respect to $G_F$ and $h_t$, the mass matrix for the neutral CP even
Higgs bosons is given by
\begin{eqnarray} {\cal M}^2 &=&
  m_A^2 
  \left( \begin{array}{cc}
      s_\beta^2         & -s_\beta c_\beta \\
      -s_\beta c_\beta  & c_\beta^2 
          \end{array}  \right)  
+ m_Z^2 
  \left( \begin{array}{cc}
      c_\beta^2         & -s_\beta c_\beta \\
      -s_\beta c_\beta  & s_\beta^2 
          \end{array}  \right)  
+ \del 
  \left( \begin{array}{cc}
      0  & 0 \\ 0 & 1 
          \end{array}  \right),  \nonumber \\
\del &=& \frac{3 m_t^4 G_F}{\sqrt{2}\pi^2}
           \frac{1}{s_\beta^2}
 \left[  \log \frac{\mx^2}{m_t^2}
       + \frac{A_t^2}{\mx^2} \left( 1 - \frac{A_t^2}{12 \mx^2} \right)
 \right].
\label{eq:delta}
\end{eqnarray} 
Here $s_\beta,c_\beta$ denote $\sin\beta,\cos\beta$. The bottom Yukawa
coupling as well as the higgsino mass parameter have been neglected
($\mu \ll \mx$).  The orthogonal diagonalization of this mass
matrix defines the CP even mixing angle $\alpha$. Only three
parameters govern the Higgs sector: the pseudo-scalar Higgs mass,
$m_A$, $\tan\beta$, and $\del$, which describes the corrections arising
from the supersymmetric top sector. For the scan of SUSY parameter
space we will concentrate on two particular values of the trilinear
mixing term, $A_t=0$ and $A_t=\sqrt{6}\mx$, which commonly are
referred to as no mixing and maximal mixing. \medskip 

\begin{figure}[ht] 
\begin{center}
\includegraphics[height=8.0cm]{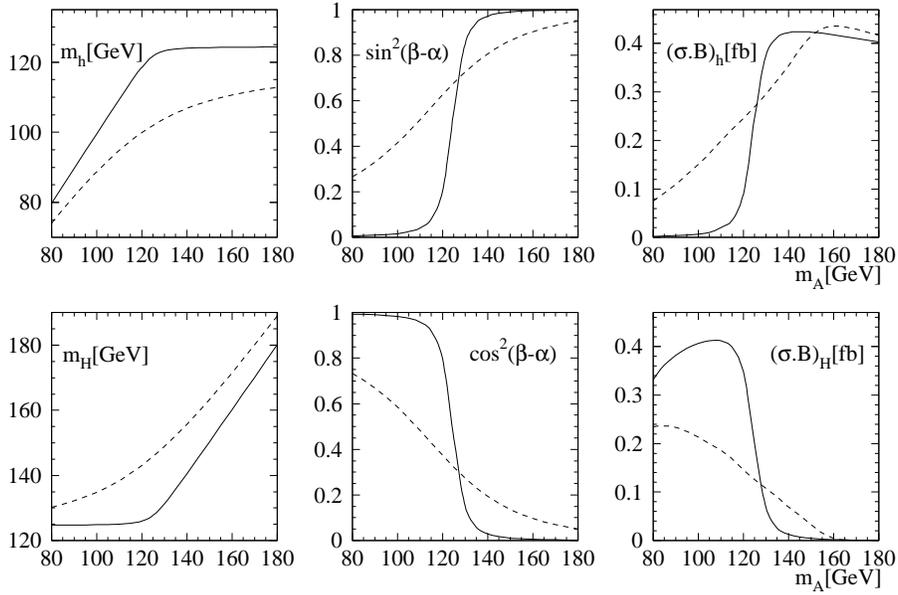} \vspace*{.1cm}
\caption[]{\label{fg:para} 
Variation of Higgs boson masses, couplings
to gauge bosons, and signal rate, $\sigma\cdot B(\tau\tau)$, for the
CP even MSSM Higgs bosons as a function of the pseudoscalar Higgs
mass. The complementarity of the search for the lighter $h$ (upper
row) and heavier $H$ (lower row) is shown for $\tan\beta=4,30$
(dashed, solid lines). Other MSSM parameters are fixed to
$\mu=200$~GeV, $\mx=1$~TeV, and maximal mixing.}  
\end{center} 
\end{figure}

Varying the pseudoscalar Higgs boson mass, one finds saturation for
very large and very small values of $m_A$ -- either $m_h$ or $m_H$
approach a plateau:
\begin{alignat}{5} m_h^2 &\simeq m_Z^2
(c_\beta^2-s_\beta^2)^2 + s_\beta^2 \del 
&&\qquad \text{for} \quad m_A \rightarrow \infty \notag \\ 
m_H^2 &\simeq m_Z^2 + s_\beta^2 \del
&&\qquad \text{for} \quad m_A \rightarrow 0.
\end{alignat} 
For large values of $\tan\beta$ these plateaus meet at
$m_{h,H}^2\approx m_Z^2+\del$. Smaller $\tan\beta$ values decrease the
asymptotic mass values and soften the transition region between the
plateau behavior and the linear dependence of the scalar Higgs masses
on $m_A$. These effects are shown in Fig.~\ref{fg:para}, where the
variation of $m_h$ and $m_H$ with $m_A$ is shown for
$\tan\beta=4,30$.  The small $\tan\beta$ region will be
constrained by the LEP2 analysis of $Zh,ZH$ associated production,
essentially imposing lower bounds on $\tan\beta$ if no signal is
observed.\footnote{Although the search for MSSM Higgs bosons at the
Tevatron is promising~\cite{tev} we only quote the $Zh,ZH$ analysis of
LEP2~\cite{lep} which is complementary to the LHC processes under
consideration. The LEP2 reach is estimated by scaling the current
limits for ${\cal L}=158$~pb$^{-1}$ and $\sqrt{s}=189$~GeV~\cite{lep}
to ${\cal L}=100$~pb$^{-1}$ and $\sqrt{s}=200$~GeV.}

The theoretical upper limit on the light Higgs boson mass, to two loop
order, depends predominantly on the mixing parameter $A_t$, the
higgsino mass parameter $\mu$ and the soft-breaking stop mass
parameters, which we treat as being identical to a supersymmetry
breaking mass scale: $m_Q=m_U=\mx$~\cite{one_loop}. As shown in
Fig.~\ref{fg:para}, the plateau mass value hardly exceeds
$\sim$130~GeV, even for large values of $\tan\beta$, $\mx=1$~TeV, 
and maximal mixing ~\cite{two_loop}. Theoretical limits
arising from the current LEP and Tevatron squark search as well as the
expected results from $Zh,ZH$ production at LEP2 assure that the
lowest plateau masses are well separated from the $Z$ mass
peak. \medskip

% --------------------------------------------------------

The production of the CP even Higgs bosons in weak boson fusion is
governed by the $hWW,HWW$ couplings, which, compared to the SM case,
are suppressed by factors $\sin(\beta-\alpha),\cos(\beta-\alpha)$,
respectively~\cite{mssm}. In the $m_h$ plateau region (large $m_A$),
the mixing angle approaches $\alpha=\beta-\pi/2$, whereas in the $m_H$
plateau region (small $m_A$) one finds $\alpha\approx-\beta$.  This
yields asymptotic MSSM coupling factors of unity for $h$ production
and $|\cos(2\beta)|\gtrsim 0.8$ for the $H$ channel, assuming
$\tan\beta \gtrsim 3$. As a result, the production cross section of
the plateau states in weak boson fusion is essentially of SM strength.
In Fig.~\ref{fg:para} the SUSY suppression factors for $\sigma(qq\to
qqh/H)$, as compared to a SM Higgs boson of equal mass, are shown as a
function of $m_A$. The weak boson fusion cross section is sizable
mainly in the plateau regions, and here the $h$ or $H$ masses are in
the interesting range where decays into $\bar bb$ and $\tau^+\tau^-$
are expected to dominate. \smallskip

Crucial for the observability of a Higgs boson are the $\tau\tau$
or $bb$ couplings of the two resonances. Splitting the couplings into
the SM prediction and a SUSY factor, they can be written as
\begin{alignat}{7}
h_{bbh}&= \frac{m_b}{v}
\left( -\frac{\sin\alpha}{\cos\beta}\right) 
&&= \frac{m_b}{v}
\Bigg( \sin(\beta-\alpha) - \tan\beta \; \cos(\beta-\alpha) \Bigg),
\notag \\
h_{bbH}&= \frac{m_b}{v}
\quad \; \frac{\cos\alpha}{\cos\beta} 
&&= \frac{m_b}{v}
\Bigg( \cos(\beta-\alpha) + \tan\beta \; \sin(\beta-\alpha) \Bigg)
\label{eq:h_bbh} 
\end{alignat} 
and analogously for the $\tau$ couplings. Since for effective
production of $h$ and $H$ by weak boson fusion we need
$\sin^2(\beta-\alpha)\approx 1$ and $\cos^2(\beta-\alpha)\approx 1$,
respectively, the coupling of the observable resonance to $\bar bb$
and $\tau\tau$ is essentially of SM strength. The SUSY factors for the
top and charm couplings are obtained by replacing $\tan\beta\to
-1/\tan\beta$ in the final expressions above. They are not enhanced
for $\tan\beta>1$. This leads to $\bar bb$ and $\tau\tau$ branching
ratios very similar to the SM results. In fact, in the plateau regions
they somewhat exceed the SM branching ratios for a given
mass. \medskip

\begin{figure}[hb] 
\begin{center} 
\includegraphics[height=8.0cm]{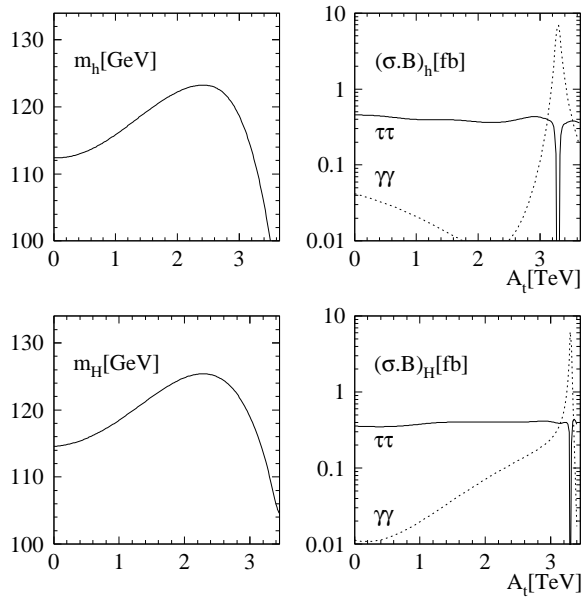} \vspace*{.1cm}
\caption[]{\label{fg:sig}
Mass of the CP even Higgs bosons and weak boson fusion rates
$\sigma\cdot B(\tau\tau,\gamma\gamma)$ as a function of the trilinear
mixing term, $A_t$. Curves are shown for $\mx=1$~TeV and $\mu=400$~GeV
with $m_A=130$~GeV, $\tan\beta=30$ ($h$: upper row), and
$m_A=105$~GeV, $\tan\beta=22$ ($H$: lower row).}
\end{center} 
\end{figure}

The $\tau\tau h$ and $\tau\tau H$ couplings vanish for $\sin\alpha=0$
and $\cos\alpha=0$, respectively, or $\sin(2\alpha)=0$. In leading
order, as well as in the simple $\del$-approximation given in
eq.(\ref{eq:delta}), this only happens in the unphysical limits
$\tan\beta=0,\infty$. Including further off-diagonal contributions to
the Higgs mass matrix might introduce a new parameter region for the
mixing angle $\alpha$: the off-diagonal element of the Higgs mass
matrix and thereby $\sin(2\alpha)$ can pass zero at finite $m_A$ and
$\tan\beta$. Indeed, by also considering the dominant contribution with
respect to $(\mu/\mx)$, one finds~\cite{one_loop}
\begin{alignat}{5} 
\left({\cal M}^2\right)_{12} &=
 -m_A^2 s_\beta c_\beta 
 -m_Z^2 s_\beta c_\beta 
 \left[ 1 - \frac{\tan\beta}{8\pi^2} \; \frac{h_t^4}{g^2} \; 
            \frac{\mu A_t^3}{\mx^4} \right], \notag \\ 
\sin(2\alpha) &=
2 \; \frac{\left({\cal M}^2\right)_{12}}{m_H^2-m_h^2} \; ,
\end{alignat} 
and $\sin(2\alpha)$ may vanish in the physical region. The exact
trajectory $\sin(2\alpha)=0$ in parameter space depends strongly on
the approximation made in perturbative expansion; we observe this
behavior for large $A_t \gtrsim 3\mx$, \ie in part of the non-mSUGRA
parameter space. If the observed Higgs sector turns out to be located
in this parameter region, the vanishing coupling to $bb,\tau\tau$
would render the total widths small. This can dramatically increase
the $h/H\to\gamma\gamma$ branching ratio, even though
$\Gamma(h/H\to\gamma\gamma)$ may be suppressed compared to the SM
case. This situation is shown in Fig.~\ref{fg:sig}, where the scalar
masses and the $\tau\tau$ and $\gamma\gamma$ rates are shown as a
function of $A_t$: the vanishing of the $\tau\tau$ rate is associated
with a very large increase of $\sigma B(\gamma\gamma)$. Note that the
variation of Higgs masses and decay properties with $A_t$ is quite
mild in general, apart from this $\sin(2\alpha)=0$ effect.

% --------------------------------------------------------

% ==================================================================
\section{Higgs Search in Weak Boson Fusion}\label{sec:three}
 
Methods for the isolation of a SM Higgs boson signal in the weak boson
fusion process ($qq\to qqh,qqH$ and crossing related processes) have
been analyzed for the $H\to\gamma\gamma$ channel~\cite{rz} and for
$H\to\tau\tau$~\cite{rzh}. The analysis for the MSSM is completely
analogous: backgrounds are identical to the SM case and the changes
for the signal, given by the SUSY factors for production cross
sections and decay rates, have been discussed in the previous section.

For the $h,H\to\gamma\gamma$ signal, the backgrounds considered are
$\gamma\gamma jj$ production from QCD and electroweak processes, and
via double parton scattering~\cite{rz}. It was found that the
backgrounds can be reduced to a level well below that of the signal,
by tagging the two forward jets arising from the scattered
(anti)quarks in weak boson scattering, and by exploiting the excellent
$\gamma\gamma$ invariant mass resolution expected for the LHC
detectors~\cite{CMS,ATLAS}, of order 1~GeV.

For $h,H\to\tau\tau$ decays, only the semileptonic decay channel of
the $\tau$ leptons, $\tau\tau\to\ell^\pm h^\mp \sla \pt$ is
considered, assuming the $\tau$-identification efficiencies and
procedures described by ATLAS for the inclusive $H,A\to\tau\tau$
search~\cite{rzh,ATLAS}. According to the ATLAS study, hadronic $\tau$
decays, producing a $\tau$ jet of $\et>40$~GeV, can be identified with
an acceptance of 26\% while rejecting hadronic jets with an efficiency
of 99.75\%. In weak boson fusion, and with the $\tau$ identification
requirements of Refs.~\cite{rzh,ATLAS} which ask for substantial
transverse momenta of the charged $\tau$ decay products
($\pt(\ell^\pm)>20$~GeV and $\pt(h^\mp)>40$~GeV), the Higgs boson is
produced at high $\pt$. In the collinear $\tau$ decay approximation,
this allows reconstruction of the $\tau^\pm$ momenta from the
directions of the decay products and the two measured components of
the missing transverse momentum vector~\cite{ATLAS,tautaumass}. Thus,
the Higgs boson mass can be reconstructed in the $\tau\tau$ mode, with
a mass resolution of order 10\%, which provides for substantial
background reduction as long as the Higgs resonance is not too close
to the $Z\to\tau\tau$ peak.

With these $\tau$-identification criteria, and by using double forward
jet tagging cuts similar to the $h,H\to\gamma\gamma$ study, the
backgrounds can be reduced below the signal level, for SM Higgs boson
masses between 105 to 150~GeV and within a 20~GeV invariant mass
bin. Here, irreducible backgrounds from `$Zjj$ events' with subsequent
decay of the (virtual) $Z,\gamma$ into $\tau$ pairs, as well as
reducible backgrounds with isolated hard leptons from $Wj+jj$ and
$b\bar bjj$ events, have been considered. Moreover, it was shown that
a further background reduction, to a level of about 10\% of the
signal, can be achieved by a veto on additional central jets of
$\et>20$~GeV between the two tagging jets. This final cut makes use of
the different gluon radiation patterns in the signal, which proceeds
via color singlet exchange in the $t$-channel, and in the QCD
backgrounds, which prefer to emit additional partons in the central
region~\cite{bjgap,bpz_minijet}.

Using the SUSY factors of the last section for production cross
sections and decay rates, one can directly translate the SM results
into a discovery reach for supersymmetric Higgs bosons. The expected
signal rates, $\sigma B(h/H\to\tau\tau,\gamma\gamma)$ are shown in
Figs.~\ref{fg:para},\ref{fg:sig}. They can be compared to SM rates,
within cuts, of $\sigma B(H\to\tau\tau)=0.35$~fb and $\sigma
B(H\to\gamma\gamma)=2$~fb for $m_H=120$~GeV. Except for the small
parameter region where the $\tau\tau$ signal vanishes, and for very
large values of $m_A$ (the decoupling limit), the $\gamma\gamma$
channel is not expected to be useful for the MSSM Higgs search in weak
boson fusion. The $\tau\tau$ signal, on the other hand, compares
favorably with the SM expectation over wide regions of parameter
space. The SUSY factors for the production process determine the
structure of $\sigma\cdot B(h/H\to\tau\tau)$. Apart from the typical
flat behavior in the asymptotic plateau regions they strongly depend
on $\beta$, in particular in the transition region, where all three
neutral Higgs bosons have similar masses and where mixing effects are
most pronounced.

% --------------------------------------------------------
\begin{figure}[ht] 
\begin{center} 
\includegraphics[width=8.0cm]{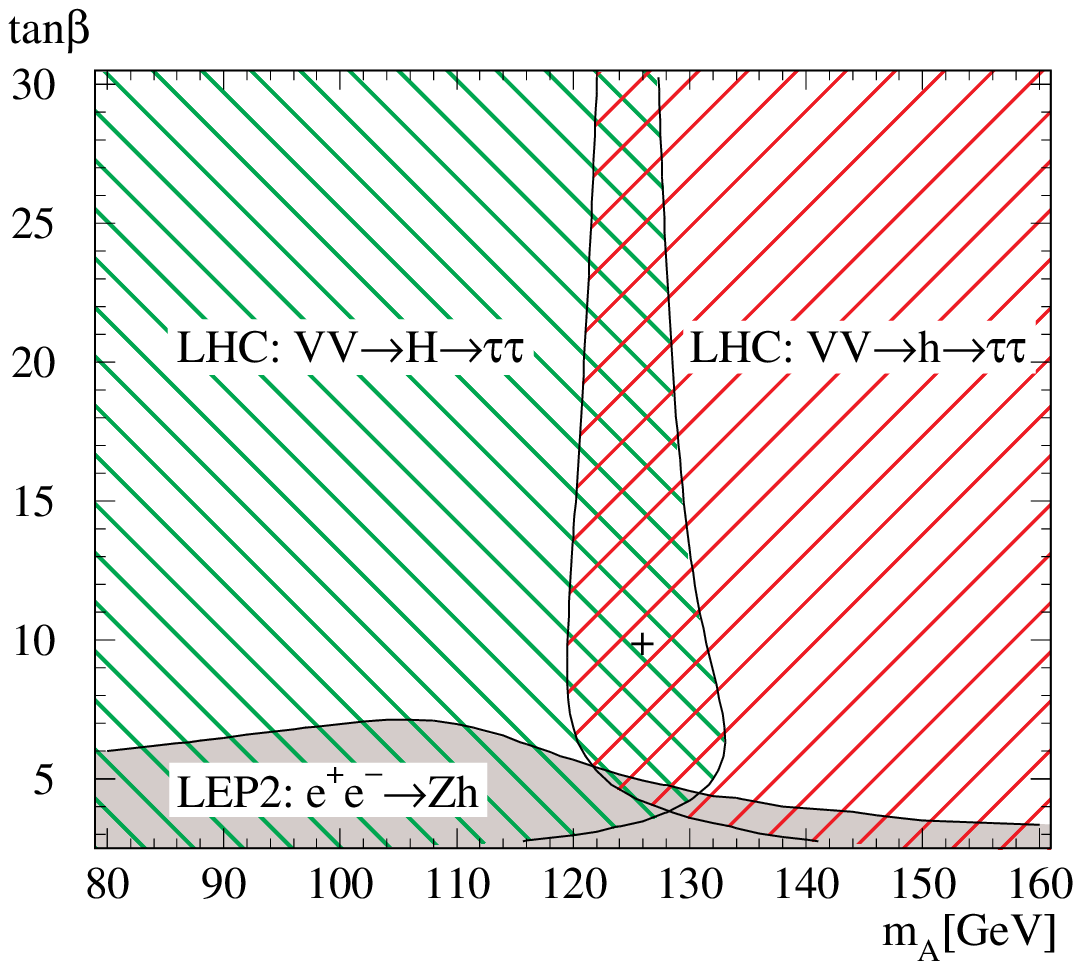} \hspace*{0.4cm}
\includegraphics[width=8.0cm]{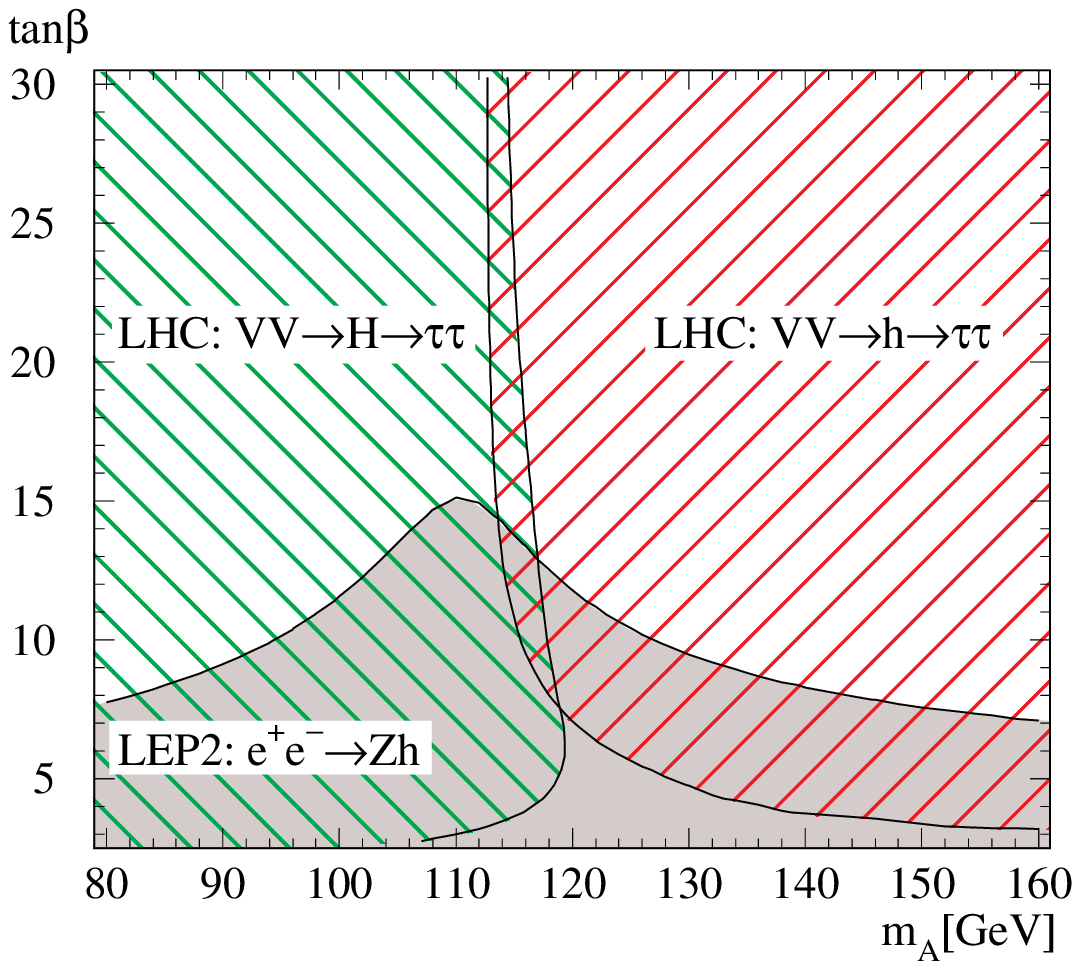}  \vspace*{.1cm}
\caption[]{\label{fg:plane}
$5\sigma$ discovery contours for $h\to\tau\tau$ and $H\to\tau\tau$ in
weak boson fusion at the LHC, with $100$~fb$^{-1}$. Also shown are the
projected LEP2 exclusion limits (see text). Results are shown for SUSY
parameters as in Fig.~\ref{fg:para}, for maximal mixing (left) and no
mixing (right). The marked point is illustrated in
Fig.~\ref{fg:mass}.}
\end{center} 
\end{figure}

\begin{figure}[ht] 
\begin{center} 
\includegraphics[width=7cm,angle=90]{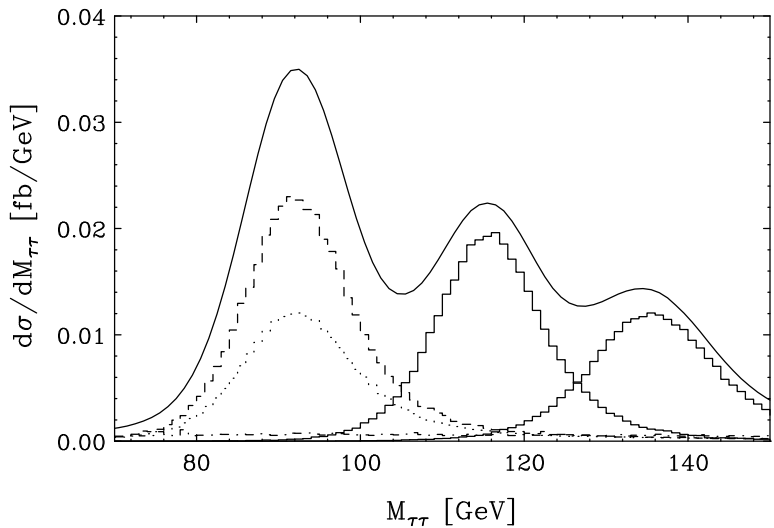} \vspace*{.1cm}
\caption[]{\label{fg:mass}
Expected $\tau$ pair invariant mass distribution for the signal (solid
histograms) and backgrounds for the search described in the text and
MSSM parameters marked in Fig.~\ref{fg:plane}.  Individual background
curves correspond to QCD $Zjj$ (dashed) and electroweak $Zjj$ (dotted)
production, and to the combined $Wj+jj$ and $b\bar{b}jj$ reducible
backgrounds (dash-dotted).  The sum of signal and backgrounds is shown
as the solid line. The three peaks correspond to $Z,h$, and $H$
production.}
\end{center} 
\end{figure}

Given the background rates determined in Ref.~\cite{rzh}, which are of
order 0.03~fb in a 20~GeV mass bin, except in the vicinity of the
$Z$-peak, the expected significance of the $h/H\to \tau\tau$ signal
can be determined. 5~$\sigma$ contours for an integrated luminosity of
100~fb$^{-1}$ are shown in Fig.~\ref{fg:plane}, as a function of
$\tan\beta$ and $m_A$. Here the significances are determined from the
Poisson probabilities of background fluctuations~\cite{rzh}.  Weak
boson fusion, followed by decay to $\tau$-pairs, provides for a highly
significant signal of at least one of the CP even Higgs bosons. Even
in the low $\tan\beta$ region, where LEP2 would discover the light
Higgs boson, the weak boson fusion process at the LHC will give
additional information. Most interesting is the transition region,
where both $h$ and $H$ may be light enough to be observed via their
$\tau\tau$ decay. A possible $\tau\tau$ invariant mass spectrum for
this scenario, with backgrounds, is shown in Fig.~\ref{fg:mass}. The
observation of a triple peak, corresponding to $Z$, $h$ and $H$ decays
to $\tau\tau$, requires very specific SUSY parameters, of
course. Fig.~\ref{fg:mass} illustrates the cleanness of the weak
boson fusion signal, however.

% --------------------------------------------------------

% ==================================================================
\section{Summary}\label{sec:four}

We have shown that the production of CP even MSSM Higgs bosons in weak
boson fusion and subsequent decay to $\tau$ pairs gives a significant
($>5\sigma$) signal at the LHC. This search, with
$\lesssim$100~fb${}^{-1}$ of integrated luminosity, and supplemented
by the search for $h/H\to\gamma\gamma$ in weak boson fusion, should
cover the entire MSSM parameter space left after an unsuccessful LEP2
search, with a significant overlap of LEP2 and LHC search regions. The
two CERN searches combined provide a no-lose strategy by themselves for 
seeing a MSSM Higgs boson. At the very least, the weak boson fusion
measurements provide valuable additional information on Higgs boson
couplings.

Our analysis here and in Ref.~\cite{rzh} should be considered as a
proof of principle, not as an estimate of the ultimate sensitivity of
the LHC experiments. A variety of possible improvements need to be
analyzed further.

\begin{itemize}

\item[--] For a Higgs resonance close to the $Z$ peak ($m_h\lesssim
110$~GeV) a shape analysis is needed to estimate the significance of
the Higgs contribution. Our sensitivity estimates are solely based on
event counting in a 20~GeV invariant mass bin.

\item[--] A trigger on the forward jets in weak boson fusion events
might allow a reduction of the transverse momentum requirement for the
$\tau$ decay lepton. A lower lepton $\pt$ threshold would
significantly increase the signal rate.

\item[--] The $\tau$ identification criteria and the rejection of the
$\bar bb$ background has been optimized for the inclusive
$A/H\to\tau\tau$ search~\cite{ATLAS}, not for the weak boson fusion
events considered here. Because of the lower backgrounds to the weak
boson fusion process, some of the requirements can be relaxed, leading
to a larger signal rate.

\item[--] Our analysis is based on parton level simulations. A full
parton-shower analysis, including hadronization and detector effects,
should be performed to optimize the cuts, and to assess efficiencies.

\end{itemize}

The present analysis relies only on the typical mixing behavior of the
CP even mass eigenstates, and on the observability of a SM Higgs
boson, of mass up to $\sim$150~GeV, in weak boson fusion. This
suggests that the search discussed here might also cover an extended
Higgs sector as well as somewhat higher plateau masses, \eg for very
large squark soft-breaking mass parameters. Because decays into $\tau$
pairs are tied to the dominant decay channel of the intermediate mass
range Higgs boson, $h/H\to \bar bb$, the search for a $\tau\tau$
signal in weak boson fusion is robust and expected to give a clear
Higgs signal in a wide class of models.

%%%%%%%%%%%%%%%%%%%%  ACKNOWLEDGMENTS  %%%%%%%%%%%%%%%%%%%%

\acknowledgements

This research was supported in part by the University of Wisconsin
Research Committee with funds granted by the Wisconsin Alumni Research
Foundation and in part by the U.~S.~Department of Energy under
Contract No.~DE-FG02-95ER40896.

%%%%%%%%%%%%%%%%%%%%%%%  REFERENCES  %%%%%%%%%%%%%%%%%%%%%%%

\bibliographystyle{plain}

\end{document}